\documentclass[preprint,showpacs,preprintnumbers,amsmath,amssymb]{revtex4-1}
\usepackage{graphicx}
\usepackage{bm}
\usepackage{color}
\usepackage{amssymb}
\usepackage{epsfig}
\usepackage{amsmath}
\begin{document}
\title{
Extended states with Poisson spectral statistics}
\author{Triparna Mondal, Suchetana Sadhukhan, Pragya Shukla}
\affiliation{Department of Physics, Indian Institute of Technology, Kharagpur, India}
\date{\today}

%\widetext

\begin{abstract} 	

Contrary to prevailing notion we find that the spectrum associated with the extended states in a complex system may  belong to the Poisson universality class if the system is subjected to a specific set of constraints. Our results are based on an exact theoretical as well as numerical analysis of column constrained chiral ensembles with circulant off-diagonal blocks and are relevant   for a complete understanding of  the eigenfunction localization and related physical properties.

\end{abstract}

%\pacs{  PACS numbers: 05.40.-a, 05.30.Rt, 05.10.-a, 89.20.-a}
                                                 
\maketitle

Statistical behavior of the eigenvalues  and eigenfunctions of linear operators play an important role in characterizing the complex systems e.g. their universalities, critical point behavior and phase transitions \cite{gmw, mj, psijmp}. Based on extensive studies, the level repulsion is generally believed to be associated with the delocalized wave-dynamics; the statistics in this  case can be modeled by one of the universality classes of stationary random matrix ensembles \cite{mj}. The other extreme, that is, the localized dynamics is characterized by a lack of level-repulsion, implying uncorrelated eigenvalues and Poisson spectral-statistics. The connectivity between eigenstates dynamics and spectral statistics is not confined only to extreme cases but 
is extended to  partially localized states too \cite{mj}.  For example, the eigenstates at metal-insulator transition are known to display the multifractal behavior alongwith a scale-invariant spectral statistics, with fractal dimension related to two-point spectral correlations \cite{mj, krav}.
Recent studies of the manybody localization also indicate the significance of spectral statistics as a criteria for  varying degree of eigenfunction localization e.g. distinguishing between ergodic  and non-ergodic extended states (with spectral statistics modeled by Wigner-Dyson ensembles and Rosenzweig-Porter ensembles respectively) \cite{krav, psrp}.
A proper understanding of this connection is therefore  highly desirable but, as the present work indicates,  seems still incomplete. This is because, contrary to typical cases, 
here we present one example in which the extended eigenstates go hand-in hand 
with Poisson spectral statistics.
Our work is based on an exact theoretical analysis of an ensemble of  chiral matrices with circulant off-diagonal blocks with random entries; keeping in view the atypical aspect of the results, we numerically verify them for two types of disorder. 
(This also motivated us for an anxious search of other similar studies which led us to  \cite{bir}).

Ensembles of chiral matrices have turned out to be good models for statistical behavior of a wide range of complex systems \cite{gmw, slng, chal, garcia} e.g Hamiltonians of bipartite lattice structures with two interconnected sublattices \cite{evan}.  Circulant matrices, a special case of Toeplitz matrices  appear in many areas too e.g. Hamiltonians for topological lattice structures represented in site-basis, statistical signal processing and information-theory \cite{dp, dgk, go}.  A combination of two matrix types i.e chiral matrix with circulant off-diagonal blocks can therefore serve as a model e.g. for a bipartite lattice with two topological sublattices.  
A knowledge of their exceptional statistical behavior can therefore provide important information about the physics of a wide range of complex systems.

A generic $2N \times 2N$ chiral matrix $H$ can be described as
\begin{eqnarray}
H= \left( {\begin{array}{cc}   0  & C  \\  C^{\dagger} & 0 \end{array} } \right).
\label{ch1}
\end{eqnarray}
where $C$ is in general $N\times(N+\nu)$ real, complex or quaternion matrix, (based on the nature of exact anti-unitary symmetry of $H$).   
%The eigenvalues of $H$ occur in pairs of $\pm \lambda$ or are zero and 
The statistical behavior of the eigenvalues/eigenfunctions of $H$-matrix  depends on the nature of $C$-matrix. 
%In absence of any other constraints except chiral and anti-unitary symmetries,  
For example, if $C$ belongs to a Hermitian matrix subjected to no  other constraint except  same strength for almost all elements,  its bulk spectral as well as eigenfunction correlations can then be modeled by the Wigner-Dyson universality class.   The chiral symmetry however  
induces an additional level repulsion near zero eigenvalue which results in different spectral correlations near the origin and away from the  bulk. The behavior of $C$ is however expected to change if additional constraints are imposed. Dictated by the feasibility of theoretical analysis, we choose $C$ to be a $N \times N$ real, circulant matrix; it is defined by the condition  (later referred as the circulant  constraint)
\begin{eqnarray}
C_{kl} = c_{(k-l) \; mod \; N} 
\label{cir1}
\end{eqnarray}.

The properties of circulant matrices are well-known and can easily be derived \cite{dp}.  
Defining $\Lambda$ as the eigenvalue matrix  of $C$ (with entries $\Lambda_{jl} =\lambda_l \; \delta_{jl}$) and $U$ as a $N \times N$ matrix with its columns as the eigenvectors $U_m$, $m=1,\ldots, N$, the matrix $C$ can be expressed as  $C=U \Lambda U^{\dagger}$.
%\begin{eqnarray}C=U \lambda U^*
%\end{eqnarray}
An eigenvector $U_m$ of a circulant matrix $C$ can be given as
\begin{eqnarray}
U_m= {1\over \sqrt{N}} \left(1, \omega_m, \omega_m^2, \ldots, \omega_m^{N-1} \right)^{T} 
\label{um}
\end{eqnarray}
with $\omega_m = {\rm e}^{2\pi i m /N}$  and $m=1, 2, \ldots, N$. The  corresponding eigenvalue is
\begin{eqnarray}
\lambda_m =  \sum_{k=0}^{N-1} c_{N-k} \; \omega_m^k
\label{lam1}
\end{eqnarray} 
%As clear from the above, $\lambda_N =  \sum_{k=0}^{N-1} c_k$. 
where $c_{N} \equiv c_0$. For real $c_k$,  the eigenvalues  satisfy $M$ pairs-wise relations (with $N=2 M+1$ for $N$ odd and $N=2M+2$ for $N$ even) : $\lambda_{N-k} = \lambda^*_{k}$ except for $\lambda_N$ which does not form any pair: $\lambda_N = \sum_{k=0}^{N-1} c_k$. For $N$ even,  $\lambda_{N/2}$ is unpaired too ($\lambda_{N/2}= \sum_{k=0}^{N-1} \; c_{N-k} \; (-1)^k$).
As clear from eq.(\ref{um}),  all circulant matrices share the same eigenvectors ($U_m$ does not depend on the matrix elements of $C$) and all eigenvectors are extended in the basis space: the inverse participation ratio (IPR), a standard measure for the eigenstate localization, defined as   $I_{2}(U_m)= \sum_{k=1}^N \mid U_{km} \mid^4$ for each $U_m$, with $m =1 \to N$, turns out to be $1/N$.    

As the definition (\ref{cir1}) indicates, each row (column) of a circulant matrix $C$ is a cyclic shift of the row (column) above it; $C$ therefore depends on $N$ free parameters $c_k$, $k=0,1,\ldots,N-1$.  Further imposing the constraint 
\begin{eqnarray}
\sum_{k=0}^{N-1} c_k = \alpha 
\label{circ}
\end{eqnarray}
with $\alpha$ as a real constant,  makes $C$ a special case of a row (column) constrained matrix:
%its columns and rows satisfying similar constraints: 
\begin{eqnarray}
\sum_{l=1}^{N} C_{kl}=\sum_{l=1}^{N}C_{lk}=\alpha.
\label{chcc1}
\end{eqnarray}
Here $\alpha$, being same for each column or row, will hereafter be referred as the column (or row) constant.   
As discussed in \cite{ss1},  the column sum rule on the entries of a matrix  manifests itself in form of constraints on its eigenfunctions and eigenvalues. For $C$, these can be given as 
\begin{eqnarray}
%${\rm Det } \;\left( H-\alpha I  \right) = 0$,
\sum_{k=1}^{N}U_{kn}=0 \qquad \; n < N, \qquad   
\sum_{k=1}^{N}U_{kN}= \sqrt{N}.
\label{eu}
\end{eqnarray}
It is easy to check that the above condition is satisfied by eq.(\ref{um}).   The circulant constraint in presence of  column constraints leads to the condition 
 \begin{eqnarray}
 \lambda_N = \alpha
\label{cons1}
\end{eqnarray}
The above can be seen by using eq.(\ref{chcc1}) for first row or column ($k=1$) alongwith eq.(\ref{um}) and the relation $ \sum_{l=1}^N \;  {\rm e}^{2\pi i n  (l-1)/N} =N \; \delta_{Nn}$
which gives $\sum_{n=1}^{N-1}  \sum_{l=1}^N \; \lambda_n \; \omega_n^{l-1} = 0$.

The combination of circulant and column constraints put new conditions on the sum of eigenvalues. Eqs.(\ref{cir1}, \ref{um}) alongwith eigenvalue equation for $C$ give
\begin{eqnarray}
\sum_{n=1}^N \lambda_n \;  \omega_n^{k-l} = N \; c_{(k-l) \; mod  \; N}.
\label{cons1}
\end{eqnarray}
Further  using eq.(\ref{chcc1}) for first row or column ($k=1$) alongwith eq.(\ref{um}) gives 
$\sum_{n=1}^N  \sum_{l=1}^N \; \lambda_n \; \omega_n^{l-1} = N \alpha$. But as $ \sum_{l=1}^N \;  {\rm e}^{2\pi i n  (l-1)/N} =N \; \delta_{Nn}$,  this again gives $\lambda_N = \alpha$.

As $H$ is  a chiral matrix,  its eigenvalues  and eigenvectors can be expressed in terms of  those of $C$. Let $E$ be the eigenvalue matrix ($E_{mn} =e_n \delta_{mn}$) and $O$ as the eigenvector matrix of $H$, with $O_{kn}$ as the $k^{th}$ component of the eigenvector $O_n$ corresponding to eigenvalue $e_n$. 
Now consider an eigenvector $U_n$ of $C$ corresponding to the eigenvalue $\lambda_n$: $C \; U_n = \lambda_n U_n$ and $C^{\dagger} \; U_n = \lambda_n^* \; U_n$ (being circulant matrices, both  $C, C^{\dagger}$ have same set of eigenvectors). This in turn implies $U_n$ as the eigenvector of 
$C C^{\dagger} = C^{\dagger} C$ with eigenvalue $|\lambda_n|^2$. Further as $\lambda_n = \lambda_{N-n}^*$ for $n < N$, this gives
\begin{eqnarray}
C C^{\dagger} \; \left(U_n +U_{N-n} \right)= C^{\dagger} C \;  \left(U_n +U_{N-n} \right) &=& |\lambda_n|^2 \; \left(U_n +U_{N-n} \right),  \qquad  n < N \\
C C^{\dagger} \; U_N = C^{\dagger} C \; U_N &=& |\lambda_N|^2 \; U_N .
\label{un}
\end{eqnarray} 
%Due to chiral symmetry, the  eigenvalues, say $e_n$, of $H$ exist in pairs and can be expressed 
The above alongwith  eq.(\ref{ch1}) implies that the  eigenvalues  of $H$ exist in equal and opposite pairs; let us refer such pairs as ${e}_n, e_{n+N}$ with $e_n=\mid \lambda_n\mid, e_{n+N} =-\mid \lambda_n\mid$, $1\le n \le N$.  The eigenvector pair $O_n, O_{n+N}$ corresponding to eigenvalue pair $e_n, e_{n+N}=\pm |\lambda_n|$ can in general be written as $\left(\begin{array}{cc} X_n  \\ \pm Y_n \end{array}\right)$.  Eq.(\ref{ch1}) then gives $C \; Y_n = \lambda_n \; X_n$ and $C^{\dagger} \;  X_n = \lambda_n \; Y_n$ which leads to $C^{\dagger} C \; Y_n = |\lambda_n|^2 \; Y_n$ and $C C^{\dagger} \; X_n = |\lambda_n|^2 \; X_n$. A comparison with eq.(\ref{un}) then implies 
\begin{eqnarray}
X_n = Y_n = \eta \; \left(U_n + (1-\delta_{nN}) \; U_{N-n} \right).
\label{on}
\end{eqnarray}  
where the real constant $\eta$ can be determined by orthogonality condition on $O_n$: $\eta= 1/2$ for $n \not= N, N/2$ and $\eta= 1/\sqrt{2}$ for $n=N, N/2$ (with $n=N/2$ case applicable for $N$ even).
The above alongwith eq.(\ref{eu}) leads to the conditions $\sum_{k=1}^{2N} O_{kn}=0$ for the eigenvector $O_n$ corresponding to an eigenvalue $e_n\not=\lambda_N$  and $\sum_{k=1}^{2N} O_{kn}=\sqrt{2 N}$ if $e_n = \lambda_N$. Note these conditions are indeed consistent with  the column constrained nature of $H$-matrix which follows from the similar nature of its blocks $C$.  The above also gives the inverse participation ratio  (IPR) for an eigenvector $O_n$ of $H$ as (for $\alpha \not=0$)
\begin{eqnarray}
 I_2(O_n) = {3 \over 4N},  \; \; {\rm for} \; n \not=N, 2N, \qquad I_2(O_N)=I_2(O_{2N}) ={1\over 2N}
\label{i2} 
\end{eqnarray}
Note however, for the case $\alpha=0$, $\lambda_N=0$ which leads to degenerate pair $e_N, e_{2N}=0$ with corresponding   eigenvectors as $\left(\begin{array}{cc} U_N  \\ 0 \end{array}\right)$ and $\left(\begin{array}{cc} 0  \\  U_N \end{array}\right)$. As a consequence for the case $\alpha=0$,
\begin{eqnarray}
I_2(O_N) =I_2(O_{2N}) = {1\over N}, \qquad {\rm with} \qquad I_2(O_n) = {3 \over 4N},  \; \; {\rm for} \; n \not=N, 2N.
\label{i20}
\end{eqnarray}

Our next step it to consider the ensemble density $\rho(H)$ of  $H$ which can subsequently be used to derive the joint probability distribution (JPDF) $P(E, O)$ of its eigenvalues and eigenfunctions. Following from eq.(\ref{ch1}), 
\begin{eqnarray}
\rho(H) = J(H | C) \; \rho_c(C)
\label{rhoh}
\end{eqnarray}
 with $J(H|C)$ as the Jacobian of transformation from $C$-space to $H$-space and $\rho_c(C)$ as the ensemble density of $C$. Following maximum entropy hypothesis, the system can be described by the distribution $\rho_c(C)$ that maximizes Shannon's information entropy
$I[\rho_c(C)] = - \int \rho_c(C)\; {\rm ln} \rho_c(C)\; {\rm d}\mu(C)$
under known constraints. Due to circulant constraint along with column/row constraint, $C$ has only $N-1$ free parameters and  $\rho_c(C)$ depends on the distribution of only $N-1$ matrix elements in any one of the rows or columns. For example, the first two moments of the $C$-entries in the first row are subjected to the constraint 
$\sum_{l=1}^{N} \langle C_{1l} \rangle  = \sum_{n=0}^{N-1} \langle c_n \rangle =\alpha$ and  $\sum_{k, l=1}^{N} \langle C_{1k} \; C_{1l} \rangle =\sum_{k, l=0}^{N-1} \langle c_{k} \; c_{l} \rangle = \alpha^2$
which  can be combined to give $\sum_{k, j} v_{k j} = 0$
with $v$ as the $N \times N$ covariance matrix  with elements $v_{kl} \equiv \langle c_{k} \; c_{l} \rangle -\langle c_{k} \rangle \langle c_{l} \rangle $. Here $\langle . \rangle$ implies the  ensemble averaging. The maximum entropy principle then leads to 
 Gaussian form of $\rho_c$:
\begin{eqnarray}
\rho_c(C) =  \mathcal{N} \;  {\rm exp}\left[-  \sum_{k,l=2}^N  {1\over 2 v_{kl}}  \; (C_{1k} -\mu_k ) (C_{1l} -\mu_l )  \right]  \; F_c 
\label{rhoc}
\end{eqnarray}
with $\mathcal{N} $ as a normalization constant, $\mu_l  = \langle C_{1l} \rangle $ and the function $F_c$ gives the circulant as well as column/row constraint:
\begin{eqnarray}
 F_c  \equiv \delta \left(\sum_{l=1}^N  C_{1l} - \alpha \right) \; \prod_{k,l=1}^N \delta(C_{kl} - c_{(k-l) \; mod \; N}). 
\label{fc}
\end{eqnarray}

The  Gaussian form of $\rho_c$ in eq.(\ref{rhoc}) results due to the constraints on  the $1^{\rm st}$ and  $2^{nd}$ order moments of first row off-diagonals only.  
Higher order moments of the latter can be subjected to similar constraints too which would lead to non-Gaussian ensembles of chiral  column-constrained matrices. A most generic form of $\rho_c$ can be given in terms of the JPDF of circulant variables $c_j$, with $j=0 \to N-1$:
$\rho_c(C) =  \mathcal{N} \; \rho_0(C_{12}, C_{13}, \ldots, C_{1N})  \; F_c$, with $F_c$ given by eq.(\ref{fc}).

 The joint eigenvalue-eigenvector distribution $P(\lambda; U) $ of $C$ can be derived by a transformation from $C$-matrix space to $\lambda, U$-space: $P(\lambda; U) = J_c(\lambda, U|C) \; \rho_c(C)$ with $J_c$ as the Jacobian of transformation. The latter depends on the derivatives ${\partial C_{kl} \over \partial \lambda_n}, {\partial C_{kl} \over \partial U_{kn}} $ which can be derived from the relation $C_{kl}= \sum_{n=1}^N \lambda_n U_{kn} U^*_{ln}$ and the orthogonality relation of the eigenfunctions. 
%$\sum_{k,l} \left(U_{kn} U^*_{km} \right) \left(U^*_{ln} U_{lm} \right) = \delta_{nm}$. 
As  the eigenvectors for all circulant matrices are same and with constant components, $C$ varies with respect to its  eigenvalues only which leads to $J_c(\lambda, U |C) = constant$.
Note an absence of the eigenvalue-repulsion in the Jacobian alongwith $I_2(U_n) = {1 \over {N}}$ for a generic eigenvector indicates the existence of extended states with uncorrelated eigenvalues even at a single matrix level. 
%Although the disorder may introduce spectral correlations, here we consider only those cases which preserve the lack of correlations. 
 The disorder however may introduce spectral correlations. As an example, we consider the ensemble density given by eq.(\ref{rhoc}) with $\mu_k=0$, $k=1 \dots, N$ and $v_{kl} = (2  \gamma)^{-1} \delta_{kl}$. This leads to 
\begin{eqnarray}
\rho_c(C) = {\mathcal N} \; {\rm exp} \left[-\gamma \; \sum_{k=2}^N \mid C_{1k} \mid^2 \right] \;F_c = {\mathcal N} \; {\rm exp} \left[-{\gamma \over (N+1)} \; \sum_{k,l=1,k\neq l}^N  \mid C_{kl} \mid^2 \right] \;F_c 
\label{rho1}
\end{eqnarray}
The sum $S_1 \equiv   \sum_{k,l=1; k \not= l}^{N} \mid C_{kl} \mid^2 $ over  all  off-diagonal squares can be expressed in terms of $\lambda_n$ and $U_n$,

\begin{eqnarray}
S_1 &=& \frac{1}{2} \sum_{m,n=1}^N \; \sum_{k=1}^{N}  \mid \lambda_m U_{km} U^*_{kn}- \lambda_n U_{kn} U^*_{km}\mid^2
\label{a1}
\end{eqnarray}
Eq.(\ref{um}) gives $U_{km}  =\frac{1}{\sqrt N} \; \omega_m^{k-1}$ which on substitution in eq.(\ref{a1}) leads to
\begin{eqnarray}
P(\lambda ; U) =   {\mathcal N} {\rm exp}\left[- \frac{\gamma }{2 N^2(N+1)} \sum_{m,n=1}^N \; \sum_{k=1}^{N} \;
\mid \lambda_m \; \omega_{m-n}^{k-1} - \lambda_n \; \omega_{n-m}^{k-1} \mid^2  \right] \; F_c
\label{plu}
\end{eqnarray}
where the constraint function $F_c$ is now expressed in terms of the constraints on the eigenvalues and eigenfunction 
\begin{eqnarray}
 F_c  \equiv E_c \; \prod_{k,l=1}^N \delta(U_{kl} -\omega_l^{k-1}). 
\label{fc1}
\end{eqnarray}
where $E_c=\delta \left(\lambda_N - \alpha \right)  \; \prod_{k=1}^{N-1} \left(\lambda_{N-k} - \lambda^*_{k}\right)$ for $N$ odd and 
$E_c = \delta \left(\lambda_N - \alpha \right)  \; \left(\lambda_{N/2} - \sum_{k=0}^{N-1} c_{N-k}(-1)^k\right) \; \prod_{k=1; \not=N/2}^{N-1} \left(\lambda_{N-k} - \lambda^*_{k}\right)$ for $N$ even.
and $\omega_n$ are constants. 
As expected from eq.(\ref{um}), the eigenvector distribution $P_u(U)$ is non-random:
\begin{eqnarray}
P_u(U)  = \prod_{k,l=1}^N \;\delta(U_{kl} -\omega_l^{k-1}). 
\label{puu}
\end{eqnarray}
Eq.(\ref{plu}) can then be written as
$P(\lambda; U) = P_{\lambda} (\lambda) \; P_u(U)$ where $P_{\lambda} (\lambda)$ is the JPDF of the eigenvalues 
\begin{eqnarray}
 P_{\lambda} (\lambda) = {\mathcal N} {\rm exp}\left[- \frac{\gamma }{N+1} \sum_{m=1}^N \; \mid \lambda_m \mid^2 \right] \; E_c. 
\label{pll}
\end{eqnarray}
where the exponent in eq.(\ref{pll}) is obtained by  using the relation $\sum_{k=0}^{N-1} \cos \left({4 \pi k n\over N} \right)=0$ in the exponent of eq.(\ref{plu}).  As clear from the above, the  eigenvalues of $C$ are uncorrelated.

Again from eq.(\ref{on}) along with eq.(\ref{puu}), the eigenvector distribution of $H$ is also non-random and its eigenvalue distribution can  be obtained from eq.(\ref{pll}) by replacing $|\lambda_m|$ by $e_m$: 
\begin{eqnarray}
P_{e} (e_1,\ldots, e_{2N}) = {\mathcal N} {\rm exp}\left[- \frac{\gamma }{N+1} \sum_{m=1}^N \; e_m^2 \right]  \; \prod_{n=1}^N \delta \left(e_n + e_{n+N} \right) \; \delta\left(e_N -\alpha   \right)
\label{pe1}
\end{eqnarray}
This confirms the lack of correlations among the eigenvalues in the ensemble of $H$ matrices too  although all eigenvectors (i.e $O_n$, $n=1, \ldots, 2N)$ are again extended, with their IPR given by eq.(\ref{i2}).

Although the above results are based on exact analysis, these   can be reconfirmed by a direct numerical analysis of the local fluctuation measures.  We consider an ensemble of $C$-matrices with   first row elements  $C_{1 n}$, with $n = 2\to N$,  independent of each other. The other matrix-elements  are subjected to the circulant constraint as well as row and column constraint with $\alpha=0$.  To understand the role of disorder on the local-fluctuations, we consider $C_{1 n}$ subjected to two types of disorders, namely, Gaussian as well as bimodal; the ensemble density for the first type is given by eq.(\ref{rho1}) and for the second type  by 
\begin{eqnarray}
\rho_c(C) \propto \left[\prod_{l=1}^N \left( \delta(C_{1l}-1) + \delta(C_{1l}+1) \right)\right] \; F_c.
\label{cbim}
\end{eqnarray} 

\begin{figure}
\centering
\includegraphics[width=1.2\textwidth,height=1.2\textwidth]{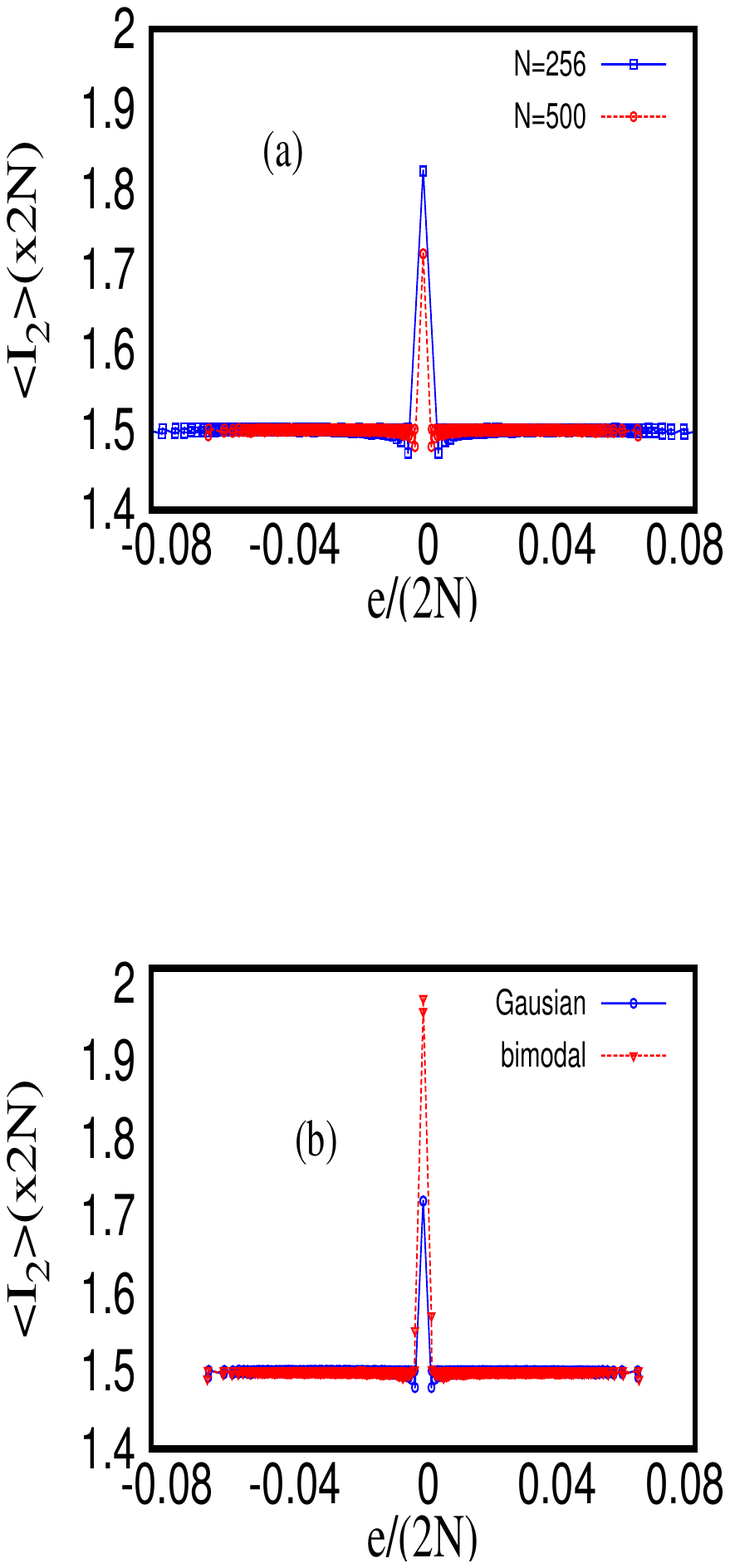} 
%\vspace*{-30 mm}
\caption{
{\bf Size and disorder dependence of ensemble averaged  inverse participation ratio $\langle I_2 \rangle$ }: 
(a) $\langle I_2 \rangle$ for two $N$-values for Gaussian disorder, (b) $\langle I_2 \rangle$ for two different types of disorders at fixed size $N=500$. The size-analogy in (a) follows on the rescaling $e \rightarrow e /(2 N)$ , $\langle I_2\rangle \rightarrow \langle I_2 \rangle\times (2 N)$. Note, as theoretically expected, $\langle I_2 \rangle$ is same for all  eigenvectors except the one at $e=0$ which corresponds to the eigenvalue $e_n=\alpha=0$. }
\label{figipr}
\end{figure}

We first consider the ensemble averaged inverse participation ratio $I_2$ for the eigenvectors $O_n$ of $H$. Theoretically eq.(\ref{um}) implies independence of  $O_n$ from the details of $C$ and eq.(\ref{i20})  characterizes their extended behavior in the basis-space. As indicated by figure \ref{figipr},  the tendency clearly survives in the presence of disorder.

\begin{figure}
\centering
\includegraphics[width=1.2\textwidth,height=1.0\textwidth]{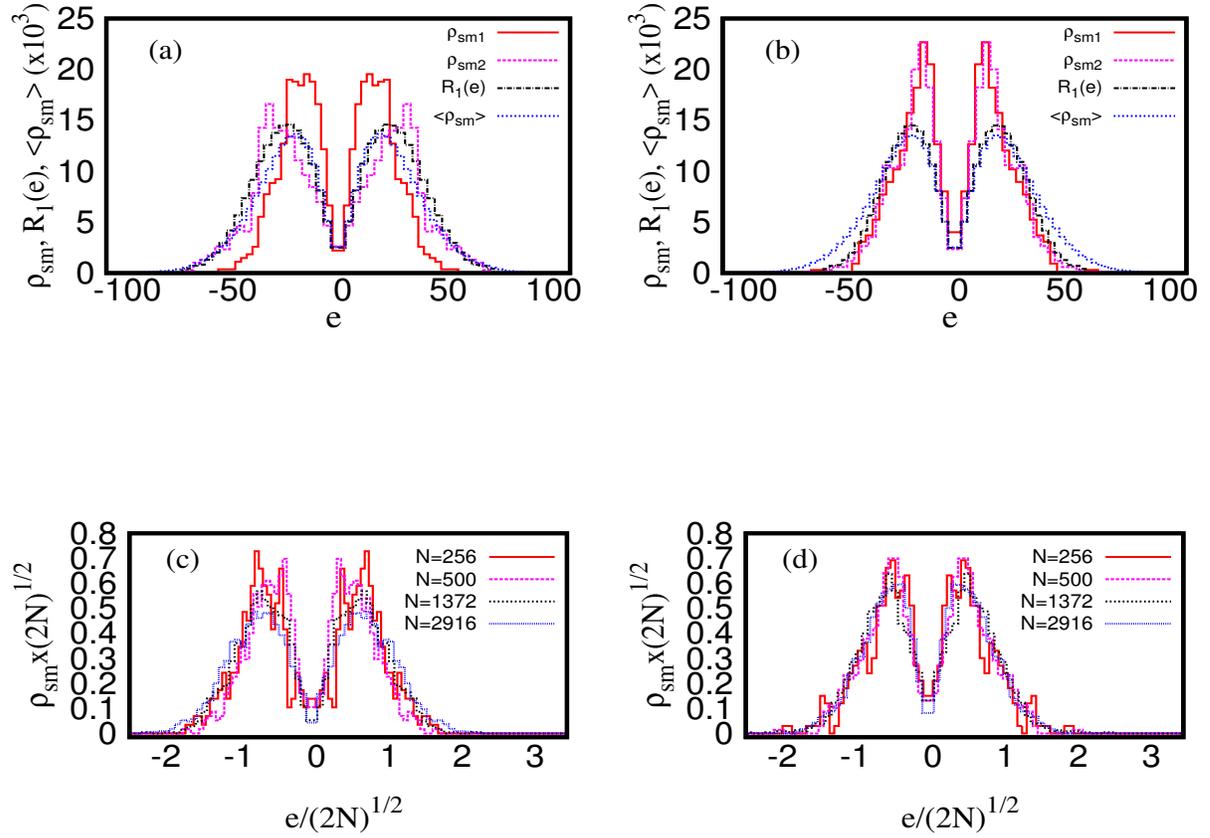} 
\caption{{\bf Non-ergodicity and size dependence of level density}:  (a)  
The figure compares the spectral averaged level-density $\rho_{sm}(e)$ for a single matrix (with $\rho_{sm1}$ and $\rho_{sm2}$ correspond to two different matrices)  to an ensemble averaged level density $R_1(e)$ as well as $\langle \rho_{sm}(e) \rangle$ for the ensemble of $H$-matrices (eq.(\ref{rhoh}))  of size $N=500$ with Gaussian disorder (eq.(\ref{rho1}),  (b) same as (a) but for bimodal disorder (eq.(\ref{cbim}), 
(c) $\rho_{sm}$ for Gaussian disordered  for different N, (with rescaling: $e \rightarrow e / \sqrt{2 N}$, $\rho_{sm} \rightarrow \rho_{sm} \times \sqrt{2 N}$).
(d) same as (c) but for bimodal disorder. Clearly  $\rho_{sm}$ fluctuates from one matrix to the other but its ensemble average also deviates from $R_1(e)$, indicating non-ergodic nature of the level density. 
}
\label{figlev}
\end{figure}

The spectral fluctuation analysis requires a prior unfolding of the eigenvalues i.e their rescaling by a locally smoothed level-density   
$\rho_{sm}(e) = {1\over 2 \Delta e} \; \int_{e-\Delta e}^{e+\Delta e} \; \rho(e) \; {\rm d}e$   with $\rho(e)=\sum_n \delta(e-e_n)$ as the level-density \cite{bg}. Assuming ergodicity, the latter can be replaced by the ensemble average $R_1(e)= \langle \rho(e) \rangle$ with $\langle . \rangle$ as an ensemble average for a fixed $e$.   
Figure \ref{figlev} compares the ensemble and the spectral average for the level density of $\rho(H)$ (eq.(\ref{rhoh}) with $\rho_c$ given by eq.(\ref{rho1}) and eq.(\ref{cbim}) for Gaussian and bimodal cases respectively) for a fixed disorder-strength. As can be seen from the figure,  the $e$-dependence of the $\rho_{sm}$ fluctuates from one matrix to the other. Further the ensemble averaged $\rho_{sm}$ deviates from $R_1(e)$ too thus indicating non-ergodic nature of the level-density \cite{bg}.
Figure \ref{figlev} also indicates   a size-dependence of the spectral averaged level-density: $\rho_{sm} \propto \sqrt{N}$ . Due to non-ergodicity of the level-density, the unfolding of the eigenvalues can be performed by the local unfolding process \cite{un}: to find  the unfolded eigenvalues, say $r_n$, the smoothed histogram of spectral density $\rho_{sm}$ for each spectra is determined and then integrated numerically, (i.e. $r_n=\int_{-\infty}^{e_N} \rho_{sm} \; {\rm d}e$).

\begin{figure}
\centering
\includegraphics[width=1.2\textwidth,height=1.0\textwidth]{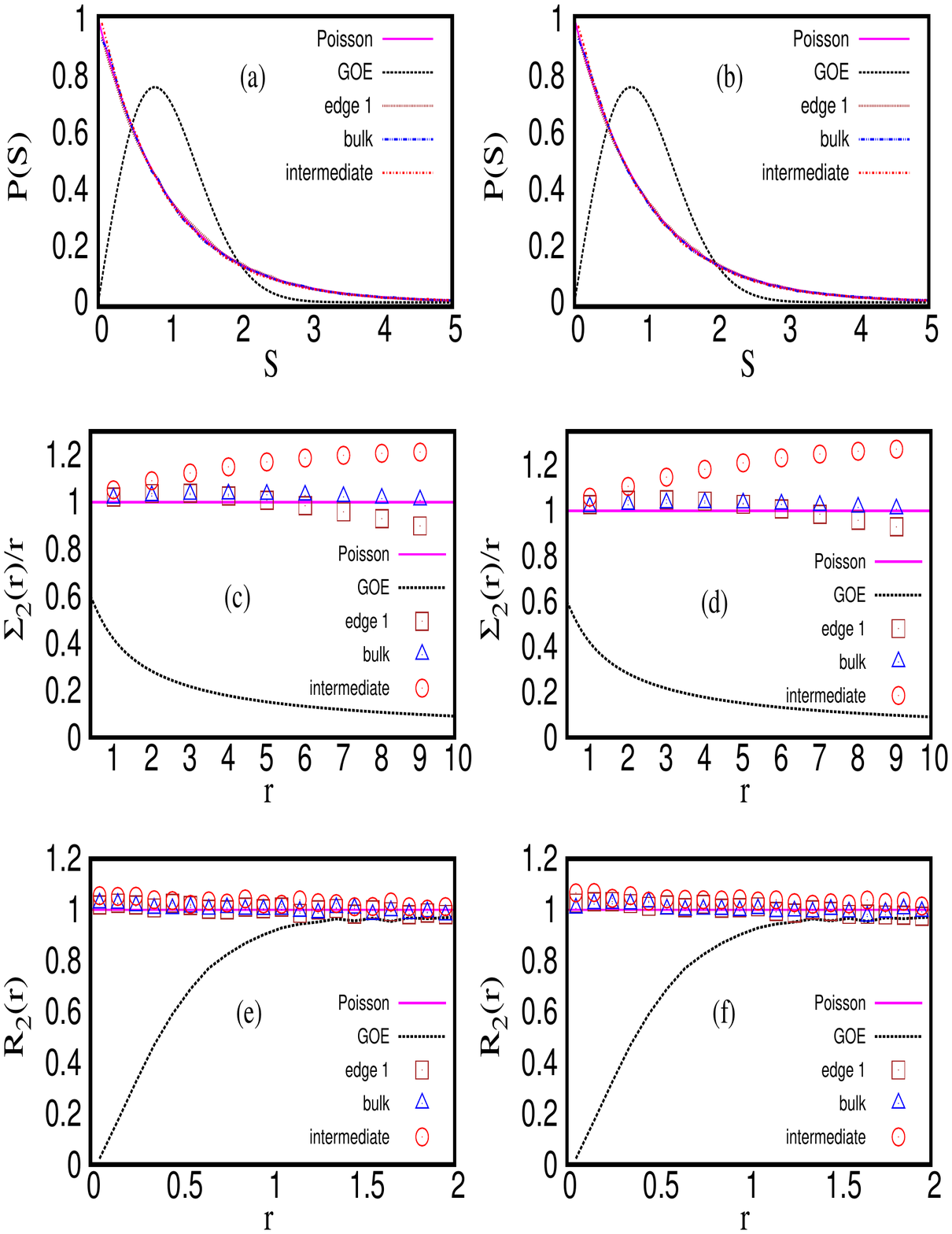}
\caption{{\bf Local spectral fluctuations for different energy regime}:
The behavior for $P(s)$, $\Sigma_2(r)$ and $R_2(r)$ for the ensemble of $H$-matrices (eq.(\ref{rhoh})  are displayed  for two disorder types for matrix size $N=2916$. The left column (i.e Fig.(a), (c), (e)) corresponds to Gaussian disorder (with $\rho_c$ given by eq.(\ref{rho1})) whereas right column (i.e Fig.(b), (d), (f)) corresponds to bimodal disorder (with $\rho_c$ given by eq.(\ref{cbim})). 
Here edge 1 refers to the region $e \sim (-1.8\pm 0.5)\times \sqrt{2N}$ ($\rho_{sm} \to 0$), bulk is $e \sim (-0.75\pm 0.05)\times \sqrt{2N}$ (maximum $\rho_{sm}$) and intermediate is $e \sim (-0.48\pm 0.06)\times \sqrt{2N}$.
The corresponding Poisson and GOE behavior for each measure are also shown. 
Clearly the behavior for both Gaussian as well as bimodal case is in good agreement with Poisson case.  }
\label{figsp}
\end{figure}

As the level-density is non-ergodic,  we consider the local fluctuations in both high and low density regions of spectrum, using only non-degenerate eigenvalues. To keep the number of levels sufficiently large for good statistics but  the mixing of different statistics minimal, we consider an optimized range $\Delta E$ ($5\% - 10\%$ of the total eigenvalues) which gives approximately $5\times 10^5$ eigenvalues for each ensemble. 
The nearest-neighbor spacing distribution $P(s)$ and the number-variance $\Sigma_2(r)$ are the standard tools for fluctuations measures for the short and long-range spectral correlations, respectively \cite{gmw, mj, psijmp}. Figure \ref{figsp} displays the   $P(s)$ and $\Sigma_2(r)$ behavior  for different energy-ranges of the spectra. The analogy with Poisson statistics for both measures reconfirms the lack of level-repulsion in the spectra. 
This uncorrelated behavior of the eigenvalues is also confirmed by the $2$-point level-density correlation $R_2(r_1,r_2)$ (the probability of finding two levels at a distance $|r_1-r_2|$). As expected on theoretical grounds,  the numerical analysis indicates the stationarity of the spectrum (indicated by almost similar local fluctuations in different spectral-range).

At this stage it is relevant to ask following question:
under what condition, the spectral statistics of the extended states can be described by the Poisson universality classes?   In the present case, the reason seems to lie in the number of independent parameters available for the dynamics in $H$ matrix-space. For $H$ as a generic Hermitian matrix, the total number of free parameters are $N(N+1)/2$. In case of a circulant matrix with column constraint, however, their number reduces to only $N-1$, resulting in $N^2-N+1$ conditions correlating matrix elements.  This in turn subjects $N$ eigenvalues and corresponding eigenvectors also to $N^2-N+1$ conditions, besides orthonormalization conditions.   The lack of free parameters can  be accommodated by the eigenvalue-eigenvector space in many ways e.g. by introducing new correlations among them or by keeping  the eigenvectors fixed and allowing only $N-1$ eigenvalues to vary as in the present case of a circulant, column-constrained matrix. As a consequence, the statistics of eigenvalues in the present case becomes independent of that of the eigenvectors, indicating a new type of basis-invariant, stationary ensemble.       

In the end, we conclude with the main insight revealed by our study: the signatures of eigenfunction dynamics on the eigenvalue statistics of complex systems are far more richer and complicated than believed so far.  The relevance of this information in a wide range of studies e.g. phase-transitions, transport properties etc. requires a better understanding of this association.

\newpage

\begin{thebibliography}{10}
\bibitem{go}
Application of these matrices in various areas can be found by web-search e.g. study of cyclic codes for error corrections, spectral estimation, watermarking, harmonic analysis, speech enhancement, interference cancellation, image restoration, sensor networks for detection, adaptive filtering, graph theory, noise reduction and blind equalization.
\end{thebibliography}
\end{document}